\documentclass[prl,twocolumn,amsfonts,amssymb]{revtex4}

\usepackage{graphicx}

\usepackage{bm}

\newcommand{\beq}{\begin{equation}}
\newcommand{\eeq}{\end{equation}}
\newcommand{\beqs}{\begin{eqnarray}}
\newcommand{\eeqs}{\end{eqnarray}}
\newcommand{\lsim}{\mathrel{\raisebox{-
.6ex}{$\stackrel{\textstyle<}{\sim}$}}}

\begin{document}

\title{Effects of Neutron-Antineutron Transitions in Neutron Stars}

\author{Itzhak Goldman$^{a,b}$, Rabindra N. Mohapatra$^c$, Shmuel Nussinov$^b$,
  and Robert Shrock$^d$}

\affiliation{(a) \ Afeka College, 6195001 Tel Aviv, Israel}

\affiliation{(b) \ Tel Aviv University, 6195001 Tel Aviv, Israel}

\affiliation{(c) \ Maryland Center for Fundamental Physics and
  Department of Physics, \\
  University of Maryland, College Park, MD 20742, USA}

\affiliation{(d) \ C. N. Yang Institute for Theoretical Physics and
  Department of Physics and Astronomy, \\
  Stony Brook University, Stony Brook, NY  11794, USA} 

\begin{abstract}

  We analyze effects of neutron-antineutron transitions in neutron stars,
  specifically on (i) cooling, (ii) rotation 
  rate, and (iii) for binary pulsars, the increase in the orbital period.
  We show that these effects are negligibly small. 

\end{abstract}

\maketitle


{\bf Introduction} \ There has long been interest in searching for
neutron-antineutron ($n$-$\bar n$) oscillations, for several
reasons. Baryon number violation (BNV) is a necessary condition for
dynamically explaining the observed baryon asymmetry in the universe
\cite{sakharov}. In addition to proton decay and BNV decays of
otherwise stably bound neutrons, which violate baryon number, $B$, as
$\Delta B=-1$ processes, another possibility is $n$-$\bar n$
transitions, which are $|\Delta B|=2$ processes.  Indeed, early on, it
was noted that $n$-$\bar n$ transitions could be relevant for the
baryon asymmetry of the universe \cite{kuzmin}.  In extensions of the
Standard Model (SM) involving an ${\rm SU}(3)_c \otimes {\rm SU}(2)_L
\otimes {\rm SU}(2)_R \otimes {\rm U}(1)_{B-L}$ gauge group (where $L$
denotes total lepton number), $n$-$\bar n$ transitions occur
naturally; furthermore, in this context, via the underlying
U(1)$_{B-L}$ gauge symmetry, $n$-$\bar n$ transitions are related to
Majorana neutrino masses that can provide an appealing explanation for
the smallness of observed neutrino masses, since the $|\Delta B|=2$
$n$-$\bar n$ operators and the $|\Delta L|=2$ Majorana neutrino mass
operators \cite{mm80,mm80b} both have $|B-L|=2$. There has thus been
continuing interest in the theory and phenomenology of possible
$n$-$\bar n$ transitions \cite{mm80}-\cite{bnv_snowmass}.  Indeed,
there are theories in which $n$-$\bar n$ transitions could be the
dominant manifestation of baryon number violation, rather than proton
decay \cite{mm80,nnb02,wise,nnblrs}.  Searches for $n$-$\bar n$
transitions have been performed using the Institut Laue-Langevin
reactor \cite{ill} and deep underground nucleon decay detectors,
including, most recently, Super-Kamiokande (Super-K) \cite{sk_nnb},
and SNO \cite{sno_nnb}, which used a limit only from $n$-$\bar n$
transitions in the deuterons in the D$_2$O.  (Limits from earlier
searches are listed in \cite{nnb_pdg}.)
  
Neutron stars have provided important tests of general relativity
\cite{ht75}-\cite{glendenning}, connections with basic nuclear physics
(e.g., \cite{jlm}-\cite{yakovlev_blanket} and references
therein), and constraints on beyond-SM (BSM) physics, in particular,
on BSM neutron interactions \cite{bgo}-\cite{berryman}. (Catalogs of
neutron stars include \cite{yakovlev_cooling,epn,atnf}.)  Much recent
work has focused on constraints on neutron-mirror neutron and dark
baryon interactions \cite{baym}-\cite{berryman}. Earlier, in
Ref. \cite{bgo}, Buccella, Gualdi, and Orlandini (BGO) analyzed the
effect of $n$-$\bar n$ transitions on the cooling of neutron stars,
and concluded that they were negligible.  Recently, Ref. \cite{fggw}
has claimed, on the contrary, that $n$-$\bar n$ transitions are very
strongly enhanced in neutron stars and that observed neutron star
properties imply an upper bound on these transitions that is much
stronger than current experimental limits and limits expected in
future experiments. For planning of the future nuclear/particle
physics program, it is crucial to confirm or refute the claims of
Ref. \cite{fggw}.  This has motivated us to reanalyze these
effects. Here we calculate the effects of $n$-$\bar n$ transitions and
subsequent $\bar n$ annihilation on (i) the cooling and (ii) rotation
rate of a neutron star and, for binary neutron-star pulsars, (iii) the
change in the orbital period.  For (i), our analysis agrees with
Ref. \cite{bgo} and decreasses the upper limit obtained there by a
factor of $4.5 \times 10^{-6}$ by using the current upper limit on
$n$-$\bar n$ transitions from terrestrial
experiments. (Ref. \cite{bgo} did not consider (ii) or (iii).) For all
of (i)-(iii), we find that $n$-$\bar n$ transitions have a negligible
effect.  Our results disagree strongly with the claims in \cite{fggw}.
  
We recall some basic properties of neutron stars.  A neutron star (NS)
arises as a remnant of a supernova explosion
(e.g. \cite{st,glendenning,raffelt}).  As compression proceeds, the
Fermi energy of degenerate electrons becomes sufficiently high that it
becomes energetically preferable for the weak reaction $e + p \to
\nu_e + n$ to take place, producing a compact object composed
predominantly of neutrons.  A typical neutron star mass is $M_{NS}
\sim 1.4 M_\odot$, where $M_\odot = 2.0 \times 10^{33}$ g is the solar
mass.  The number of neutrons in a NS of this mass is thus $N_n \sim
M_{NS}/m_n \sim 10^{57}$.  A typical NS radius is $\sim 10$ km, and
hence a typical density is $\sim 5 \times 10^{14}$ g/cm$^3$,
comparable to nuclear densities.  The stability of the neutron star
arises from a combination of neutron degeneracy pressure and the
hard-core repulsion of the neutrons. Owing to the contraction from
stellar radii to $\sim 10$ km, neutron stars have large rotation rates
with periods $P \sim 0.05 - 10$ s and large magnetic fields $B = |\vec
B| \sim 10^{12}$ Gauss.  After initially cooling mainly by neutrino
emission, subsequent long-term cooling is via photon emission. For our
analysis here, we only need a few basic inputs, as we will discuss.
With this brief sketch of relevant neutron star properties, we next
proceed to our calculations.  (We use units with $c=\hbar=1$.)


{\bf Background on $n$-$\bar n$ Transitions} \ 
  We recall some relevant background on $n$-$\bar n$ transitions.
Let us denote the basic transition amplitude as
\beq
\delta m = \langle \bar n | {\cal H}_{\rm eff} | n \rangle \ .
\label{delta_m}
\eeq
The $2 \times 2$ Hamiltonian in the $n, \bar n$ basis is then
\beq
\cal{M}=\left(\begin{array}{cc}
m_{n,{\rm eff}}  & \delta m \\
\delta m         & m_{\bar n, {\rm eff}} \end{array}\right) \ ,
\label{mat}
\eeq
where, in a nuclear medium,  
\beq
m_{n,{\rm eff}} = m_n + V_n \ , \quad
m_{\bar n, {\rm eff}} = m_n + V_{\bar n} \ .
\label{mvalues}
\eeq
Here, the nuclear potential $V_n$ is real, $V_n = V_{nR}$, but
$V_{\bar n}$ has an imaginary part representing the $\bar n N$
annihilation: $V_{\bar n} = V_{\bar n R} - i V_{\bar n I}$ (nuclear
calculations include \cite{dgr,fg,gbl,bbgr,jlm,ny,jm}).
The mixing is strongly suppressed
relative to the situation in field-free vacuum; the mixing angle goes
like
\beq
\frac{2\delta m}{|m_{n,{\rm eff}} - m_{\bar n,{\rm eff}}|} =
\frac{2\delta m}{\sqrt{(V_{n R}-V_{\bar n R})^2 + V_{\bar n I}^2}} \ll 1 \ .
\label{theta_m}
\eeq
Note that although neutron stars can have large magnetic
fields $B \sim 10^{12}$ Gauss, the energy splitting $\Delta E|_B$ due to
these magnetic fields is negligible relative to the effect of $V_I$;
\beq
|\Delta E|_B \simeq 2|\mu_n|B = 1.2 \times 10^{-5} \bigg ( 
\frac{B}{10^{12} \ {\rm G}} \bigg ) \ {\rm MeV}, 
\label{magnetic_splitting}
\eeq
where $\mu_n=-1.91[e/(2m_N)]$. 

The eigenvalues of the Hamiltonian matrix are 
\begin{widetext}
\beq
m_{1,2} = \frac{1}{2} \bigg [ m_{n,{\rm eff}} +  m_{\bar n,{\rm eff}} \pm
  \sqrt{ (m_{n,{\rm eff}} -
    m_{\bar n,{\rm eff}})^2 + 4(\delta m)^2 } \ \bigg ] \ .
\label{m1m2}
\eeq
\end{widetext}
Expanding $m_1$ for the mostly $n$ mass eigenstate
$|n_1\rangle \simeq |n\rangle$, we have
\beq
m_1 \simeq m_n + V_n - i \frac{(\delta m)^2 \, V_{\bar n I}}
{(V_{n R}-V_{\bar n R})^2 + V_{\bar n I}^2} \ .
\label{m1_taylor}
\eeq
The imaginary part represents the resultant matter instability
via annihilation of $\bar n$ with neighboring nucleons, with a rate
\beq
\Gamma_m = \frac{1}{\tau_m} = \frac{2(\delta m)^2 |V_{\bar n I}|}
      {(V_{n R} - V_{\bar n R})^2 + V_{\bar n I}^2} \ .
\label{gamma_m}      
\eeq
Hence,  $\tau_m \propto (\delta m)^{-2} = \tau_{n \bar n}^2$, where
$\tau_{n \bar n} = 1/|\delta m|$ is the time scale characterizing
$n$-$\bar n$ transitions in (field-free) vacuum. Writing
\beq
\tau_m = R \, \tau_{n \bar n}^2 \ ,
\label{ttrel}
\eeq
one has $R \sim 10^{23}$ s$^{-1}$, depending on the nuclear
medium \cite{dgr,fg,gbl,bbgr}.  

The lower bound on $\tau_{n \bar n}$ from $n - \bar n$ searches with
neutron beams from reactor is $\tau_{n \bar n} > 0.86 \times 10^8$ s
\cite{ill}.  The best lower bound on $\tau_{n \bar n}$ is from the
SuperKamiokande experiment, namely $\tau_m > 3.6 \times 10^{32}$ yr
(90 \% CL) \cite{sk_nnb}. With $R = 0.52 \times 10^{23} \ {\rm s}^{-1}
= 34$ MeV \cite{dgr,fg,sk_nnb}, this corresponds to
\beq
\tau_{n \bar n} > 4.7 \times 10^8 \ {\rm s} \ (90 \% \ {\rm CL}) \ .
\label{sk_nnb}
\eeq
The SNO experiment reported two lower limits depending on the
statistical analysis method, the more stringent of which was $\tau_m >
1.48 \times 10^{31}$ yr (90 \% CL), which, with $R = 0.248 \times
10^{23}$ s$^{-1}$ \cite{dgr,fg}, yielded $\tau_{n \bar n} > 1.37
\times 10^8$ s \cite{sno_nnb}.


{\bf Effect on Neutron Star Cooling} \ Here we study the effects of
$n$ - $\bar n$ transitions on the long-term cooling of a neutron star
(NS). Given a nonzero transition matrix element $\delta m$, there is a
finite probability $P_{n \to \bar n}(t) = |\langle \bar n |
n(t)\rangle|^2$ that a state $|n(t)\rangle$ that is initially a
neutron, $n$, at time $t=0$ will be an antineutron, $\bar n$, at time
$t$.  This will annihilate with a neighboring neutron, yielding mainly
pions (with average multiplicity $\sim 5$) and thereby depositing
energy $2m_n$. (Here, for the purposes of our estimates, we can
neglect the real parts $V_{nR}$ and $V_{\bar nR}$ in the effective $n$
and $\bar n$ masses (cf. Eq. (\ref{mvalues})), although, of course, we
do not neglect the imaginary part $V_{\bar nI}$ of $m_{\bar n,{\rm
    eff}}$.)  These pions will undergo strong reactions with adjacent
neutrons on a time scale $\sim 10^{-23}$ s, including $\pi^+ n \to
\pi^0 p$.  The $\pi^0$s produced directly from the $\bar n$
annihilation and via this charge-exchange reaction will then decay via
$\pi^0 \to \gamma\gamma$.  Energy escaping via neutrinos from $\pi^-
\to \mu^- \bar\nu_\mu$ is expected to be negligible, and its presence
would only strengthen our conclusions, since it would reduce the
energy deposition contributing to photons and hence to the NS
luminosity. The matter instability due to $n-\bar n$ transitions and
consequent annihilation is characterized by the matter decay rate
$\Gamma_m = 1/\tau_m$ given in Eq. (\ref{gamma_m}).

Using $N_n(t) = N_n(0)e^{-t/\tau_m}$ and taking into account that the
age of the universe, $t_U = 1.38 \times 10^{10}$ yrs, so $t_U \ll
\tau_m$, it follows that, to very good accuracy,
\beq
\frac{dN_n}{dt} = -\frac{N_n(0)}{\tau_m} e^{-t/\tau_m} =
-\frac{N_n(0)}{\tau_m} \ ,
\label{dndt}
\eeq
where $N_n(0)$ denotes the initial number of neutrons in the neutron
star.  Hence, the number of neutrons that transform to $\bar n$,
$N_{n \to \bar n}$, divided by the initial number of neutrons,
$N_n(0)$, is
\beq
\frac{N_{n \to \bar n}(t)}{N_n(0)} = 
\frac{1}{N_n(0)}\Big |\frac{dN_n}{dt}\Big | t = 
\bigg ( \frac{t}{\tau_m} \bigg ) < 2.8 \times 10^{-29}
\Big ( \frac{t}{10^4 \ {\rm yr}} \Big ) \ .
\label{ntonbar}
\eeq
Here we have taken $10^4$ yr as a reference time; ages of neutron
stars in the recent compendium in Ref. \cite{yakovlev_cooling} (see
also the catalogs \cite{epn,atnf}) range from roughly $t \sim 10^3$ yr to
$t \sim 10^6$ yr.

The energy deposition rate resulting from the $n$-$\bar n$ transitions
followed by annihilation is
\beqs
&& \frac{dU}{dt} = \Big | \frac{dN_n}{dt} \Big | (2m_n)
= \bigg (\frac{N_n(0)}{\tau_m} \bigg ) (2m_n) =
\frac{2M_{NS}}{\tau_m}  \ . \cr\cr
&&
\label{dudt_gen}
\eeqs
Note that with $N_n(0) \simeq M_{NS}/m_n$, the dependence on
$m_n$ largely divides out in this expression for
$dU/dt$. Evaluating Eq. (\ref{dudt_gen}) numerically, we find
\beqs
\frac{dU}{dt} &=& (4.4 \times 10^{14} \ {\rm erg/s})
\bigg ( \frac{M_{NS}}{1.4M_\odot} \bigg )
\bigg ( \frac{3.6 \times 10^{32} \ {\rm yr}}{\tau_m} \bigg ) \ . \cr\cr
&&
\label{dudt1}
\eeqs
Using the relation (\ref{ttrel}), this can also be expressed in terms
of the fundamental quantity $\tau_{n \bar n}$, as
\beqs
\frac{dU}{dt} &=& (4.4 \times 10^{14} \ {\rm erg/s})
\bigg ( \frac{M_{NS}}{1.4M_\odot} \bigg )
\bigg ( \frac{4.7 \times 10^8 \ {\rm s}}{\tau_{n \bar n}} \bigg )^2  \ .
\cr\cr
&&
\label{dudt2}
\eeqs
This is an upper bound on this energy deposition rate, since the
values that we have used for $\tau_m$ and $\tau_{n \bar n}$ are the
experimental lower bounds on these quantities.

The robustness of our calculation and the similar one in \cite{bgo},
can be understood from two fundamental properties of the physics,
namely localty and continuity.  First, the $n \to \bar n$ transition
and annihilation process is local, so the effect of $n$-$\bar n$
transitions in the full volume of the neutron star can be computed by
dividing that volume up into subvolumes equal to the volume of an
${}^{16}$O nucleus, relevant for the lower limit on $\tau_{n \bar n}$
set by the Super-K experiment \cite{sk_nnb}.  From the relation of the
radius $R_{\rm nuc}$ of a nucleus to the atomic number, $R_{\rm nuc}
\simeq (1.3 \ {\rm fm})A^{1/3}$, it follows that $R_{\rm nuc} \simeq
3$ fm for an ${}^{16}$O nucleus. The time associated with the
$n$-$\bar n$ annihilation in each of the subvolumes is then $R_{\rm
  nuc}/c \sim 10^{-23}$ s, whose inverse immediately yields a rough
estimate of the factor $R$ in Eq. (\ref{ttrel}), $R \sim 10^{23}$
s$^{-1}$, close to the result of the detailed calculations in
\cite{dgr,fg}.  Second, the neutron number density in a neutron star
is close to the nucleon number density in a nucleus such as
${}^{16}$O. Since the physics is a continuous function of the inputs,
it follows that the $R$ value calculated in ref. \cite{dgr,fg} should
also apply reasonably accurately to $n$-$\bar n$ transitions in a
neutron star. Indeed, the SNO experiment \cite{sno_nnb} obtained its
lower limit $\tau_m$ and hence $\tau_{n \bar n}$ from a search for
$n$-$\bar n$ transitions in the deuterons ${}^2$H in heavy water
D$_2$O. Since there is only a single neutron in ${}^2$H, there is no
issue of Fermi degeneracy in the calculation of $R$. Thus, from a
theoretical point of view, we do not see any reason for enhancement
of the $n-\bar{n}$ transition rate in a neutron star relative to
the rate in nuclei.

In order to determine if the energy deposition rate in
Eq. (\ref{dudt2}) has a significant effect on the neutron star, one
chooses an old neutron star that has undergone a long period of
cooling, since the fractional effect on the surface temperature $T_s$
is largest for the lowest $T_s$. Values of surface temperatures of
neutron stars extracted from observations are listed, e.g., in the
compendium in Ref. \cite{yakovlev_cooling,gravcor}; these are $T_s \sim 5
\times 10^5$ K to $10^6$ K, i.e., 50-100 eV.  The
corresponding thermal radiative luminosity is
\begin{widetext}
\beqs
L_{NS} &=& 4\pi R_{NS}^2 \sigma_{SB} T_s^4 = (7.1 \times 10^{32} \ {\rm erg/s})
\Big (\frac{R_{NS}}{10 \ {\rm km}} \Big )^2
\Big (\frac{T_s}{10^6 \ {\rm K}} \Big )^4 \ , \cr\cr
&&
\label{L_ns}
\eeqs
\end{widetext}
where $\sigma_{SB}=5.67 \times 10^{-5}$ erg/(s cm$^2$ K$^4$) is the
Stefan-Boltzmann constant.  The fractional change due to the $n$-$\bar
n$ transitions is thus $(dU/dt)/L_{NS} \lsim 10^{-18}$, which is
negligibly small.  This conclusion is in agreement with
\cite{bgo}. Our main updates relative to Ref. \cite{bgo} are (i) to
use the current value of the lower limit on $\tau_{n \bar n}$ and (ii)
to take advantage of the advances since 1987 in observational data and
theoretical modelling of neutron stars. For comparison,
Ref. \cite{bgo} utilized the limit then available, $\tau_{n \bar n} >
10^6$ s.  Since $dU/dt \propto \tau_{n \bar n}^{-2}$, using the
current lower limit on $\tau_{n \bar n}$ reduces the upper limit on
$dU/dt$ by the factor $[(10^6 \ {\rm s})/(4.7 \times 10^8 \ {\rm
    s})]^2 = 4.5
\times 10^{-6}$ relative to the value obtained with
the 1987 inputs in \cite{bgo}, strengthening the conclusion reached in
\cite{bgo} that $n$-$\bar n$ transitions have a negligible effect on
the neutron star.


{\bf Effect on Neutron Star Rotation} \ Given the result for $L_{NS}$
in Eq. (\ref{L_ns}), it also follows that $n$-$\bar n$ transitions
have a negligible effect on the rotation of a neutron star.  Let us
denote the rotation period as $P$, the angular frequency of rotation
as $\omega=2\pi/P$, and the moment of inertia as $I$, where, to a good
approximation, $I=(2/5)M_{NS}R_{NS}^2$. The rotation rate $\omega$
decreases (called the spin-down process), and hence $E_{\rm
  rot}=(1/2)I\omega^2$ decreases, due to the emission of energy via
magnetic dipole radiation by the neutron star pulsar.  Then $dE_{\rm
  rot}/dt = I \omega \dot \omega = - (2\pi)^2 I (\dot P)/P^3$, where
$\dot Q \equiv dQ/dt$ for a quantity $Q$. From the observed values of
$P_b$ and $\dot P_b$, one calculates a time $t_c = (1/2)P/\dot P$ that
is approximately characteristic of the age of the pulsar.  For a
typical neutron star of mass $M_{NS} \sim 1.4M_\odot$ and radius
$R_{NS} \simeq 10$ km, $I \simeq 10^{45}$ g cm$^2$. Making reference
to the compendium of neutron star properties in
\cite{yakovlev_cooling}, let us take the Vela pulsar as an
illustrative example. This pulsar has $P=0.0893$ s, $t_c=1.13 \times
10^4$ yr, and hence $\dot P = P/(2t_c) = 1.2 \times 10^{-13}$.
Consequently, $-\dot E_{\rm rot} \simeq 7 \times 10^{36}$ erg/s.  The
energy deposition rate $dU/dt$ in Eq. (\ref{dudt1}) or equivalently
(\ref{dudt2}), is $\lsim 10^{-22}$ of this spin-down energy
loss rate and therefore has a negligible effect on the spin-down
process.  We have carried out similar comparisons for a number of
other neutron stars with a range of values of $P$ and $\dot P$, with
the same conclusion.


{\bf Effect on Orbital Period of Binary Pulsars} \ 
We can also estimate the effect of $n$-$\bar n$ transitions on the
orbital period $P_b$ of neutron stars comprising binary pulsars. As in
Refs. \cite{gn_wimps,gn_nu}, we employ the Jeans relation \cite{jeans}
$\dot P_b/P_b= - 2\dot M/M$, where $P_b$ denotes the orbital
  period and $M= M_1 + M_2$ denotes the total mass of the binary
  system.  Let us consider, for example, the well-studied Taylor-Hulse
  binary pulsar system, PSR B1913+16 (also denoted PSR J1915+1606),
  which was used as a test of general relativity (GR)
  \cite{ht75,tw82,wnt2010,weisberg_huang}, \cite{will}.  For this system,
  $P_b=0.322997$ days (measured (to an accuracy of 1 part in $10^{11}$ -
  we do not show all of the significant figures). The total mass of
  this binary system is $M = 2.83 M_\odot$, and the mass decrease
  takes place in each of the binary members.  The analysis in
  Ref. \cite{weisberg_huang}, updating the earlier work in \cite{wnt2010},
  gives the intrinsic (int) value of $\dot P_b$ (after correcting for
  extrinsic effects) as
\beq
\dot P_{b,{\rm int}} = (-2.393 \pm 0.004) \times 10^{-12}
\label{pbdot}
\eeq
and cites the prediction from general
relativity for the value of $\dot P_b$ resulting from gravitational
radiation as
\beq
\dot P_{b,{\rm GR}} = (-2.40263 \pm 0.00005) \times 10^{-12} \ . 
\label{pbdot_gr}
\eeq
Comparing these, Ref. \cite{weisberg_huang} finds 
$\dot P_{b,{\rm int}}/\dot P_{b,{\rm GR}} = 0.9983 \pm 0.0016$, 
in agreement, to within the estimated accuracy, with the GR prediction.
The residual (res) $\dot P_{b,{\rm res}}$ is then
\beq
\dot P_{b,{\rm res}} \equiv \dot P_{b,{\rm int}} - \dot P_{b,{\rm GR}} =
(4.6 \pm 4.0) \times 10^{-15}  
\label{pdot_residual}
\eeq
so
\beq
\frac{\dot P_{b,{\rm res}}}{P_b}=(5.2 \pm 4.5) \times 10^{-12} \
     {\rm yr}^{-1} \ .
\label{pdot_over_p_residual}
\eeq
Substituting $M=2.83M_\odot$ in Eq. (\ref{dudt2}),
denoting this mass/energy loss as $\dot M_{n-\bar n} = -dU/dt$, and
using the Jeans relation, we have
\beqs
&& \frac{\dot P_{b,n \bar n}}{P_b} = -2\frac{\dot M_{n-\bar n}}{M} \cr\cr
&=& (1.1 \times 10^{-32} \ {\rm yr}^{-1})
\bigg ( \frac{4.7 \times 10^8 \ {\rm s}}{\tau_{n \bar n}} \bigg )^2 \ .
\label{pdot_over_p_nnbar}
\eeqs
Again, this is an upper limit, since the value used for
$\tau_{n \bar n}$ is the experimental lower limit \cite{sk_nnb}.
Thus, the increase in $\dot P_b/P_b$ due to possible $n$-$\bar n$
transitions and annihilation is a factor of about $10^{20}$ times
smaller than the observed residual $\dot P_{b,{\rm res}}/P_b$ for the
Taylor-Hulse binary pulsar.  We have also performed similar estimates
for other binary pulsar systems, with the same conclusions. This shows
that possible $n$-$\bar n$ transitions have a negligible effect on the
period of binary pulsars, just as they have a negligible effect on the
pulsar luminosities and spin-down rates.


{\bf Discussion and Conclusions} \ 
We note that our conclusions differ strongly with the claim in
\cite{fggw}. Ref. \cite{fggw} bases its claim of a huge enhancement of
$n$-$\bar n$ transitions in neutron stars on the effect of Fermi degeneracy.
But this degeneracy is described by the Fermi energy of the neutrons, 
$E_F = \frac{1}{2m_n}( 3 \pi^2 \rho_{n,{\rm num}})^{2/3}$, 
where $\rho_{n,{\rm num}}$ is the average number density of the
neutrons in a neutron star (corresponding to the average NS mass
density $\rho_n = m_n \rho_{n, {\rm num}})$. Since this number density
is similar to the number density of nucleons in an oxygen nucleon,
there is not a strong difference between degeneracy energies in
neutron stars and the oxygen nuclei that were the basis for the lower
limit on $\tau_{n \bar n}$ from Super-K \cite{sk_nnb}.

We conclude that, given present lower bounds on $n$-$\bar n$
transitions and resultant matter instability, the effects on the
cooling and change in rotation rate of neutron stars, and the change
in orbital period of neutron-star binary pulsars, are negligibly
small.  For this reason, terrestrial experiments to search for
$n$-$\bar n$ transitions continue to be worthwhile.

We thank Bhupal Dev for bringing Ref. \cite{fggw} to our attention and
to Shao-Feng Ge for some correspondence on \cite{fggw}. The research
of RS was partially supported by the National Science Foundation grant
NSF-22-10533.



\begin{thebibliography}{99}
\bibitem{gravcor} The temperature $T_s$ at the surface of the
  neutron star differs from the temperature $T_{s,\infty}$ measured by
  an distant observer due to a gravitational redshift factor:
  $T_{s,\infty}=T_s/(1+z)$, where $1+z=[1-(R_s/R_{NS})]^{-1/2}$, where
  $R_s=2GM/c^2=(2.95 \ {\rm km})( M/M_\odot)$ is the Schwarzschild
  radius.  For $M_{NS}=1.4M_\odot$ and $R_{NS}=10$ km, the redshift
  factor is $1+z=1.3$. Tabulated values of surface temperature for
  neutron stars are often given as $T_{s,\infty}$ (e.g.,
  \cite{yakovlev_cooling}).
\end{thebibliography}
\end{document}